\documentclass[12pt,preprint]{aastex}

\usepackage{natbib}
\usepackage{longtable}
\usepackage{rotate}
\usepackage{lscape}
\def\gtorder{\mathrel{\raise.3ex\hbox{$>$}\mkern-14mu
             \lower0.6ex\hbox{$\sim$}}}
\def\ltorder{\mathrel{\raise.3ex\hbox{$<$}\mkern-14mu
             \lower0.6ex\hbox{$\sim$}}}

\newcommand{\sst}{{\emph{Spitzer Space Telescope}}}

\shorttitle{SDWFS-MT-1: A Self-Obscured Luminous Supernova at $z\simeq0.2$}
\shortauthors{Koz{\l}owski et al.}

\begin{document}

\title{SDWFS-MT-1: A Self-Obscured Luminous Supernova at $z\simeq0.2$}

\author{Szymon~Koz{\l}owski\altaffilmark{1}, 
C.~S.~Kochanek\altaffilmark{1,2}, 
D.~Stern\altaffilmark{3}, 
J.~L.~Prieto\altaffilmark{4,14},
K.~Z.~Stanek\altaffilmark{1,2}, 
T.~A.~Thompson\altaffilmark{1,2},
R.~J.~Assef\altaffilmark{1}, 
A.~J.~Drake\altaffilmark{5}, 
D.~M.~Szczygie{\l}\altaffilmark{1,6},
P.~R.~Wo{\'z}niak\altaffilmark{7},
P.~Nugent\altaffilmark{8},
M.~L.~N.~Ashby\altaffilmark{9},
E.~Beshore\altaffilmark{10},
M.~J.~I.~Brown\altaffilmark{11},
Arjun~Dey\altaffilmark{12},
R.~Griffith\altaffilmark{3},
F.~Harrison\altaffilmark{5},
B.~T.~Jannuzi\altaffilmark{12},
S.~Larson\altaffilmark{10},
K.~Madsen\altaffilmark{5},
B.~Pilecki\altaffilmark{6,13},
G.~Pojma{\'n}ski\altaffilmark{6},
J.~Skowron\altaffilmark{1},
W.~T.~Vestrand\altaffilmark{7},
J.~A.~Wren\altaffilmark{7}
}

\altaffiltext{1}{Department of Astronomy, The Ohio State University, 140 West 18th Avenue, Columbus, OH 43210, USA; e-mail:  
{\tt simkoz@astronomy.ohio-state.edu}}
\altaffiltext{2}{The Center for Cosmology and Astroparticle Physics, The Ohio State University, Columbus, OH 43210, USA}
\altaffiltext{3}{Jet Propulsion Laboratory, California Institute of Technology, 4800 Oak Drive, Pasadena, CA 91109, USA}
\altaffiltext{4}{Carnegie Observatories, 813 Santa Barbara Street, Pasadena, CA 91101, USA}
\altaffiltext{5}{California Institute of Technology, 1200 E. California Blvd, Pasadena, CA 91125, USA}
\altaffiltext{6}{Warsaw University Observatory, Al. Ujazdowskie 4, 00-478 Warsaw, Poland}
\altaffiltext{7}{Los Alamos National Laboratory, ISR-1, MS-D466, Los Alamos, NM 87545, USA}
\altaffiltext{8}{Lawrence Berkeley National Laboratory, 1 Cyclotron Road, Berkeley, CA 94720, USA}
\altaffiltext{9}{Harvard-Smithsonian Center for Astrophysics, 60 Garden Street, Cambridge, MA 02138, USA}
\altaffiltext{10}{The University of Arizona, Department of Planetary Sciences, Lunar and Planetary Laboratory, 1629 E. University Blvd, Tucson, AZ 85721, USA}
\altaffiltext{11}{School of Physics, Monash University, Clayton 3800, Victoria, Australia}
\altaffiltext{12}{National Optical Astronomical Observatory, 950 North Cherry Avenue, Tucson, AZ 85719, USA}
\altaffiltext{13}{Universidad de Concepci{\'o}n, Departamento de Fisica, Casilla 160-C, Concepci{\'o}n, Chile}
\altaffiltext{14}{Hubble, Carnegie-Princeton Fellow}

\begin{abstract}
We report the discovery of a six-month-long mid-infrared transient, SDWFS-MT-1 (aka SN~2007va), in the {\it Spitzer} 
Deep, Wide-Field Survey of the NOAO Deep Wide-Field Survey Bo{\"o}tes field. 
The transient, located in a $z=0.19$ low luminosity ($M_{[4.5]}\simeq-18.6$ mag, $L/L_{\star}\simeq0.01$) 
metal-poor ($12+\log({\rm O/H})\simeq7.8$) irregular galaxy, 
peaked at a mid-infrared absolute magnitude of $M_{[4.5]}\simeq-24.2$ 
 in the $4.5 \micron$ {\it Spitzer}/IRAC band and emitted a total energy of at least $10^{51}$~ergs.
The optical emission was likely fainter than the mid-infrared, although our constraints on the optical
emission are poor because the transient peaked when the source was ``behind'' the Sun. 
The {\it Spitzer} data are consistent with emission by a modified black body with a temperature of $\sim1350$~K.
We rule out a number of scenarios for the origin of the transient such as a Galactic star, AGN activity, GRB, 
tidal disruption of a star by a black hole and gravitational lensing. The most plausible scenario is a 
supernova exploding inside a massive, optically thick circumstellar medium, composed of multiple 
shells of previously ejected material. If the proposed scenario is correct, then a significant fraction ($\sim10$\%)
of the most luminous supernova may be self-enshrouded by dust not only before but also after the supernova occurs.
The spectral energy distribution of the progenitor of such a supernova would be a slightly cooler version of $\eta$~Carina, peaking at 
20--30$\mu$m. 
\end{abstract}


\section{Introduction}

The first truly large variability survey of extragalactic sources at mid-infrared (mid-IR, 3.6--8.0\micron) wavelengths
is the {\it Spitzer} Deep, Wide-Field Survey (SDWFS; \citealt{2009ApJ...701..428A})
of the NOAO Deep Wide-Field Survey (\citealt{1999ASPC..191..111J}) Bo{\"o}tes field. 
The survey spans the years 2004-2008 with four epochs, covers 8 deg$^2$, and contains 
variability statistics for nearly half a million sources (\citealt{2010ApJ...716..530K}).
While the majority ($\sim 76$\%) of the extragalactic variable objects in the mid-IR are active galactic nuclei (AGNs),
there is room for serendipitous discoveries such as supernovae (SNe). 

In general, SDWFS was not expected to be interesting for SNe searches. First, the contrast
between SNe and their hosts is relatively poor in the mid-IR, essentially because of 
differences in the effective temperatures.  If we match the bolometric luminosities
of an SN and a typical host galaxy spectral energy distribution (SED), the SN contributes $\sim 70\%$ of the V-band
flux but only $\sim 8\%$ of the $3.6\mu$m flux.  Second, with only four epochs of
data, we expected SDWFS would add little to our understanding SNs even if one were 
detected.  Third, the expected rates are very low. 
Typical Type Ia (IIp) SN peak at $M_{[3.6]} \sim M_K \sim -18.4$ ($-18.1$) mag
(e.g., \citealt{2004AJ....128.3034K,2010MNRAS.403L..11M}), so SDWFS can detect
them in a single epoch only for redshifts $ z\ltorder 0.07$ ($0.06$).  Given
the SNe Ia rates from the SDSS-II Supernova Survey (\citealt{2010ApJ...713.1026D}) and SNe Type II rate (\citealt{2009PhRvD..79h3013H}), 
we would then expect to detect only of order one Type Ia SN and of order seven Type II SNe in the SDWFS survey even
if we could safely search for them at the $5\sigma$ detection level for a single
epoch ($[3.6]=19.1$~mag).  In practice, controlling the false positive rates for 
identifying variable sources means that the variability selection criteria must be 
significantly more conservative than a $5\sigma$ peak at the detection limit (see \citealt{2010ApJ...716..530K}), 
so we did not expect to detect any SNe in SDWFS.  

There is a subclass of Type IIn SNe that are far more luminous, reaching
$M_{[3.6]} \sim M_K \sim -20$ mag (e.g., \citealt{2003A&A...401..519M}), which we 
could detect up to $z\simeq 0.15$.
The leading theory for these SNe uses collisions between two massive shells of 
material to efficiently radiate the kinetic energy of the SN in the optical (\citealt{2007ApJ...671L..17S},
\citealt{2007Natur.450..390W}).  The fast moving SN ejecta collides
with an outer, slower moving shell on scales of order $10^2$~AU where the optical
depth is high enough to produce a well-defined photosphere but low enough for the
thermal energy from the shock to be radiated before adiabatic expansion converts
it back into kinetic energy.  But these SNe are rare, representing only $\lesssim 1/160$
of the local Type II SNe rate (\citealt{2009ApJ...690.1303M}), which more than balances the increased detection volume.

Here we investigate the nature the brightest mid-IR transient in SDWFS. Over a 6 month period, 
this $z\simeq 0.2$ source radiated $\sim 10^{51}$~ergs at a black body temperature 
of $\sim1350$~K.   In Section~\ref{sec:data} we present 
the available UV, optical and mid-IR data along with a Keck spectrum of the host galaxy. 
Then, in Section~\ref{sec:results} we consider a range of possible scenarios for
producing this source and conclude that a simple variant of the models for the 
hyper-luminous Type IIn SNe is the best explanation for this source. 
The paper is summarized in Section~\ref{sec:summary}.
Throughout this paper we use a standard $\Lambda$CDM
model with $(\Omega_{\Lambda}, \Omega_{\rm M}, \Omega_{k}) = (0.73, 0.27, 0.0)$ and $h=H_0/100=0.71$\footnote{To 
derive basic parameters of the transient and its host galaxy, we used the Cosmology Calculator 
(\citealt{2006PASP..118.1711W}); \tt http://www.astro.ucla.edu/$\sim$wright/CosmoCalc.html}.
All magnitudes are in the Vega system.

\section{Data}
\label{sec:data}

\subsection{Spitzer/IRAC Data}

\begin{figure}[t]
\centering
\includegraphics[width=16cm]{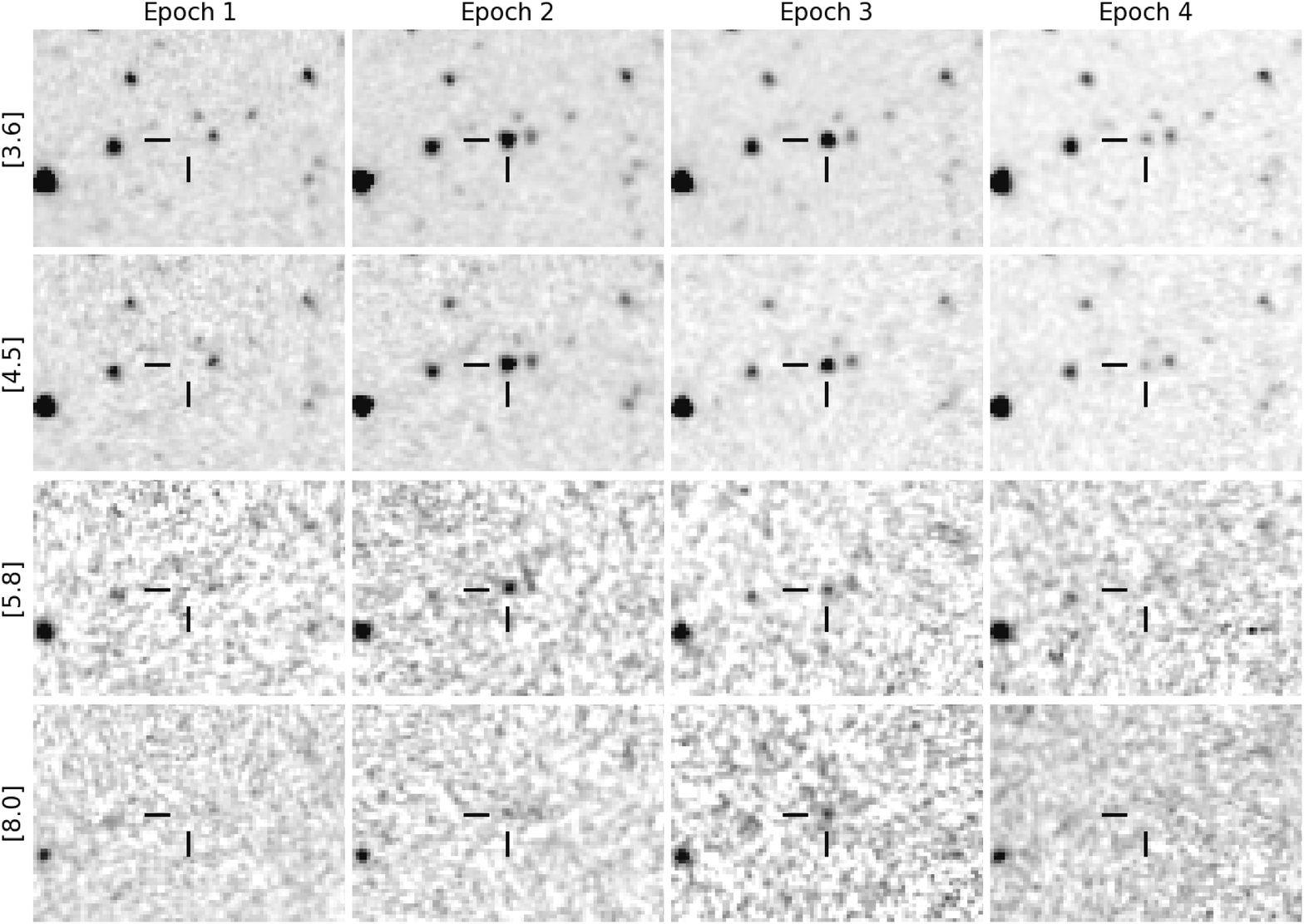}
\caption{SDWFS images, approximately $1.5\times1$ arcmin$^2$, centered on the transient. 
The rows correspond to the [3.6], [4.5], [5.8], and [8.0] bands (downward from top), 
and the columns show Epochs 1 through 4 (left to right), taken in 2004, 2007 and twice in 2008, respectively. 
The object is marked with the cross hairs. The source is clearly absent in Epoch 1 and then present in all subsequent epochs
except for the Epoch 4 [8.0] data.}                                                                                                                      
\label{fig:SDWFS}                                                                                                   
\end{figure}

SDWFS-MT-1 (aka SN~2007va; \citealt{2010CBET.2392....1K}), where MT means mid-IR transient as a parallel notation to using OT for optical transient,
is the most significantly variable source in the entire SDWFS field.  It corresponds to the 
SDWFS source SDWFS J142623.24$+$353529.1.
The [3.6] and [4.5] light curves were strongly correlated ($r=1$) and the variability amplitude 
was 76 standard deviations from the mean for its average magnitude. 
A detailed definition of these variability criteria is given in \cite{2010ApJ...716..530K}.  
The SDWFS survey consists of four epochs,
taken 3.5 years, 6 months, and 1 month apart. The object is not detected in any of the four IRAC bands in the first epoch 
(also known as the IRAC Shallow Survey, \citealt{2004ApJS..154..48E}) on 
2004 January 10-14 (${\rm JD'} = {\rm JD}-2450000 \approx 3016.5$).
The SDWFS single epoch $3\sigma$ detection limits of $[3.6]=19.67$ mag, $[4.5]=18.73$ mag, $[5.8]=16.33$ mag, and $[8.0]=15.67$ mag 
(4 arcsec apertures corrected to 24 arcsec; Vega mag), 
set our upper limits on the object brightness in the first epoch. The observed part of the transient 
peaked in the second epoch taken 2007 August 8-13 (${\rm JD'} \approx 4322.5$) at $[3.6]=15.93$ mag 
and $[4.5]=15.61$ mag (Figures~\ref{fig:SDWFS} and \ref{fig:lightcurve}). It was slightly fainter 
6 months later in epoch 3 (2008 February 2-6, ${\rm JD'} \approx 4500.5$), with  
$[3.6]=16.09$ mag and $[4.5]=15.76$ mag. Then, in a matter of a month,
the object faded by almost 2 magnitudes to $[3.6]=17.99$ mag and $[4.5]=17.72$ mag in epoch 4 on 2008 March 6-10 (${\rm JD'} \approx 4533.5$).
We present all the available {\it Spitzer} data in Table~\ref{tab:SDWFSdata}.

IRAC (\citealt{2004ApJS..154...10F}) simultaneously observes the $[3.6]/[5.8]$ or $[4.5]/[8.0]$ bands, and in
SDWFS the $[4.5]/[8.0]$ observations were taken 40 seconds after the $[3.6]/[5.8]$ ones. Each epoch consists of three 30 second 
exposures in each band, taken no more than 2 days apart, and the images for the whole mosaics were taken within 5 days. 
We confirmed that the transient was present in the individual artifact-corrected basic calibrated data (CBCD) frames 
corresponding to the position of the transient.  The mosaic image for each epoch is then a combination of three to
four of these frames, and there is no indication of problems in the SDWFS reductions (see \citealt{2009ApJ...701..428A}). 

\subsection{Keck Spectrum}
\label{sec:data_keck}

We obtained a Keck/LRIS (\citealt{1995PASP..107..375O}) spectrum of the transient's host galaxy on 2010 March 12, long after the transient was gone.
The spectrum (shown in Figure~\ref{fig:Keck_spectrum}) shows clear emission lines 
of H$\alpha$, H$\beta$, H$\gamma$, H$\delta$, H$\epsilon$, 
and H$\zeta$, [OII] at 3727\AA, [O III] at 4363\AA, 4959\AA\, and 5007\AA, [NeIII] at 3869\AA\, and 3968\AA. 
The redshift of the host galaxy is $z=0.1907$, 
so the luminosity distance is 920 Mpc, the distance modulus is 39.82 mag, and the angular scale is 3.15 kpc/arcsec.
Before measuring the line fluxes, we corrected the host galaxy spectrum for Galactic extinction (E(B$-$V$)=0.014$ mag) using \cite{1998ApJ...500..525S}.
We do not detect the [NII] 6584\AA\ line, which implies log([NII]/H$\alpha)<-1.5$. Combining it with log([OIII] 5007\AA/H$\beta)=0.78$, 
we conclude that the host is an irregular HII galaxy (see, e.g., \citealt{1991A&AS...91..285T, 2004ApJS..153..429K}).
The lack of [NII] and [NeV] 3426\AA\ lines and the clear detection of the 4363\AA~line rule 
out the presence of an AGN. 

We followed the prescription of \cite{2007ApJ...669..299P} to derive the gas-phase oxygen abundance of the host.
We measured the fluxes of the [OII] 3727\AA, [OIII] 4363\AA, 
[OIII] 4959\AA, [OIII] 5007\AA\, and H$\beta$ 4861\AA\, emission lines
to find a metallicity of $12 + \log({\rm O/H}) = 7.85\pm 0.13$.
We assumed an electron density of $n_e = 100$ cm$^{-3}$, but the
value does not change significantly with changes in the electron density (e.g.,
$12 + \log({\rm O/H}) = 7.87$ if we use $n_e = 1000$ cm$^{-3}$ instead). 
Next, we verified this metallicity measurement using the prescriptions of
\cite{2006A&A...448..955I} to get $12 + \log({\rm O/H}) = 7.79$.
As a cross check, we also used the oxygen-to-hydrogen flux ratio 
$R_{23}=({\rm [OII]+[OIII]})/{\rm H}\beta=(F_{\lambda3727}+F_{\lambda4959}+F_{\lambda5007})/F_{\lambda4861}$
method of \cite{1979MNRAS.189...95P} and
the \cite{2004MNRAS.348L..59P} O3N2 method to derive oxygen abundances.
The clear detection of the 4363\AA~line and log([NII]/H$\alpha)<-1.5$
strongly suggests that we should use the ``lower'' metallicity branch in the $R_{23}$ method.
We use the relations from \cite{1989ApJ...347..883S} and \cite{2007A&A...462..535Y}  with our measured
$\log(R_{23})=0.94$, to get oxygen abundances of $12 + \log({\rm O/H}) = 7.78$ and $7.80$, respectively.
The O3N2 method gives $12 + \log({\rm O/H})<8.0$.
All these estimates are approximately 10\% of the solar value (assuming 8.8 from \citealt{2010arXiv1005.0423D}).

\subsection{UV and Optical Data}

In order to understand the origin of this mid-IR transient, we searched for any available UV and optical data for this area of 
the sky in the relevant time span. 

The host galaxy was detected in the original NOAO Deep Wide-Field Survey (NDWFS, \citealt{1999ASPC..191..111J})
with $B_{\rm W}=23.98\pm0.05$ mag ($\lambda F_\lambda = 6.64\times 10^{41}$ ergs/s), $R=22.78\pm0.05$ mag ($1.10\times 10^{42}$ ergs/s), 
and $I=22.58\pm0.09$ mag ($8.39\times 10^{41}$ ergs/s).  It was unresolved in $1.2$ arcsec (FWHM) seeing, which
puts an upper limit on the size of the galaxy of 3.8 kpc.  In archival GALEX (\citealt{2007ApJS..173..682M})
data, the host has NUV$=24.30\pm0.17$ mag and no detection in FUV.  The GALEX source is formally 0\farcs87 from 
the SDWFS/NDWFS source, but there are no other blue sources within 20~arcsec.  Both the colors of the host and its size 
are consistent with a blue, low luminosity, low metallicity galaxy. We used the template model of an irregular galaxy 
from \cite{2010ApJ...713..970A} to estimate the mid-IR magnitudes of the host. Fitting the GALEX NUV, and 
NDWFS $B_W$, $R$, and $I$-band magnitudes of the host galaxy, and assuming $z=0.1907$, the expected apparent 
(absolute) IRAC magnitudes are
$[3.6]  = 20.87~(-18.56)$ mag,
$[4.5]  = 20.69~(-18.57)$ mag, 
$[5.8]  = 20.52~(-20.02)$ mag, and
$[8.0]  = 18.37~(-21.45)$ mag. 
This is consistent with the failure to detect the host in the first SDWFS epoch.
We estimate rest frame host luminosities of $M_B=-16.11$ mag and $M_R=-16.75$ mag,  
making this an $L/L_{\star}\simeq 0.01$ galaxy (\citealt{2005AJ....129..178K}).

Although the field was monitored by at least 5 optical variability surveys, the
peak of the transient occurred when the field was ``behind'' the Sun, so almost all
the optical data corresponds to the periods before epoch 2 and after epoch 4. 
The Bo{\"o}tes field was observed seven 
times between ${\rm JD'} =3800$ and $4620$ with the 8k$\times$8k Mosaic CCD camera on the 2.4 meter MDM 
Hiltner Telescope. The two most interesting observations took place 90 days before the second
and 84 days after the fourth SDWFS epoch, bracketing the period of the transient. In these 180 second 
exposures, we detected no flux from the transient (or the host) at a $3\sigma$ level of $R=21.32$ mag 
($4.2\times 10^{42}$ ergs/s) and $21.15$ mag ($4.9\times 10^{42}$ ergs/s), respectively.

\begin{figure}[t]
\centering
\includegraphics[width=16cm]{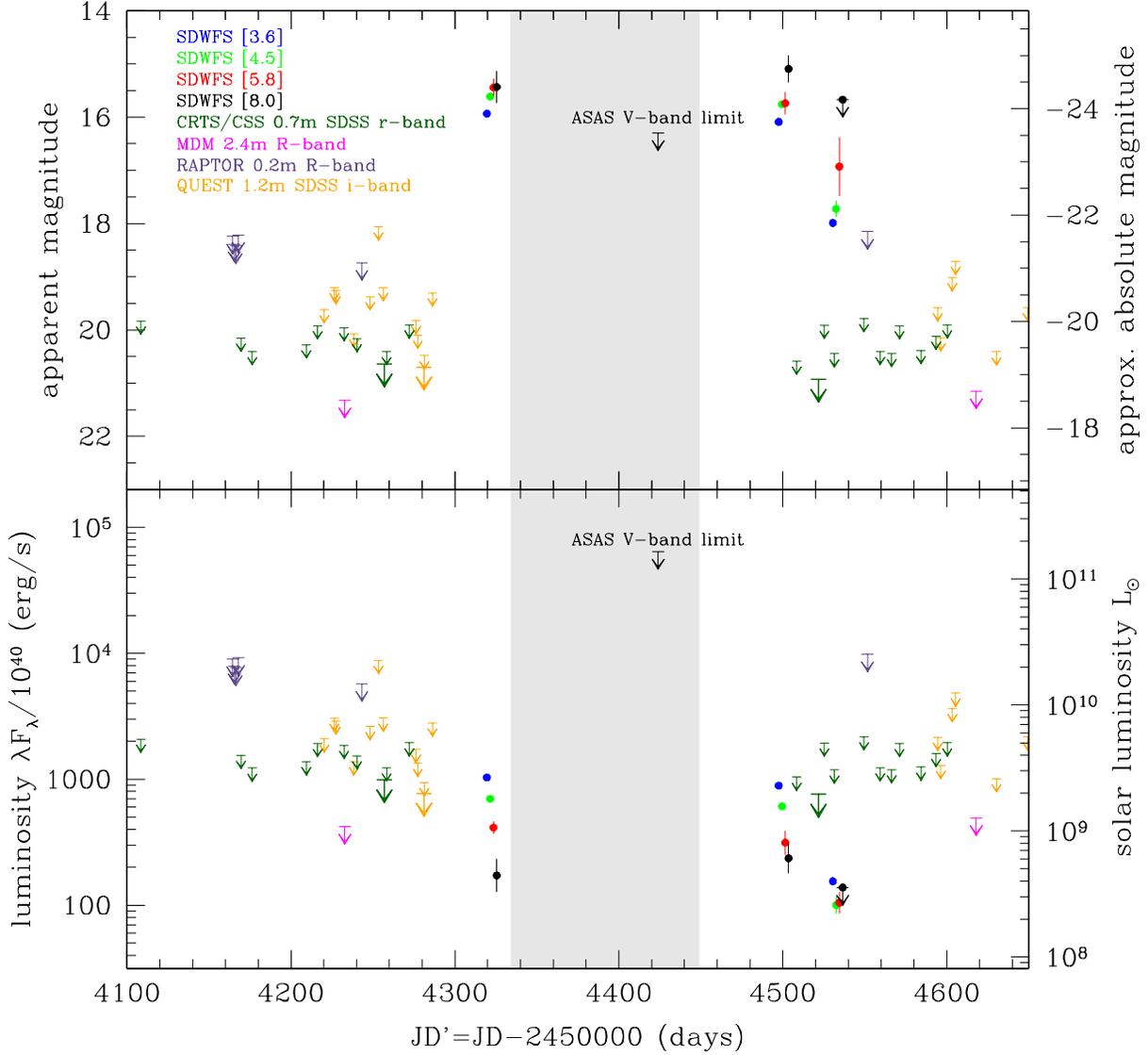}
\caption{Light curves of the transient. Top panel: observed optical and mid-IR magnitudes.
For display purposes the IRAC points are offset by $-3$ days [3.6], $-1$ day [4.5], $+1$ day [5.8] and, 
$+3$ days [8.0].  The right $y$-axis shows the absolute magnitudes without $K$-corrections.
A normal Type II SN would peak at $M \simeq -20$ mag at 3.6\micron. 
The gray area marks the region when the mid-IR transient was not visible to airmass $<1.5$. 
The large dark green symbols mark the CRTS upper limits found by stacking the three nearest epochs (small symbols) 
into one image. 
The RAPTOR upper limits (purple symbols) are stacks of 21--62 images.
The QUEST upper limits (orange) are derived based on 1--4 stacked images per epoch. 
The large symbol for QUEST combines the four nearest epochs.
Bottom panel: points from the upper panel converted to luminosity as $\lambda F_{\lambda}$.}     
\label{fig:lightcurve}                                                                                                   
\end{figure}

\begin{figure}[t]
\centering
\includegraphics[width=16cm]{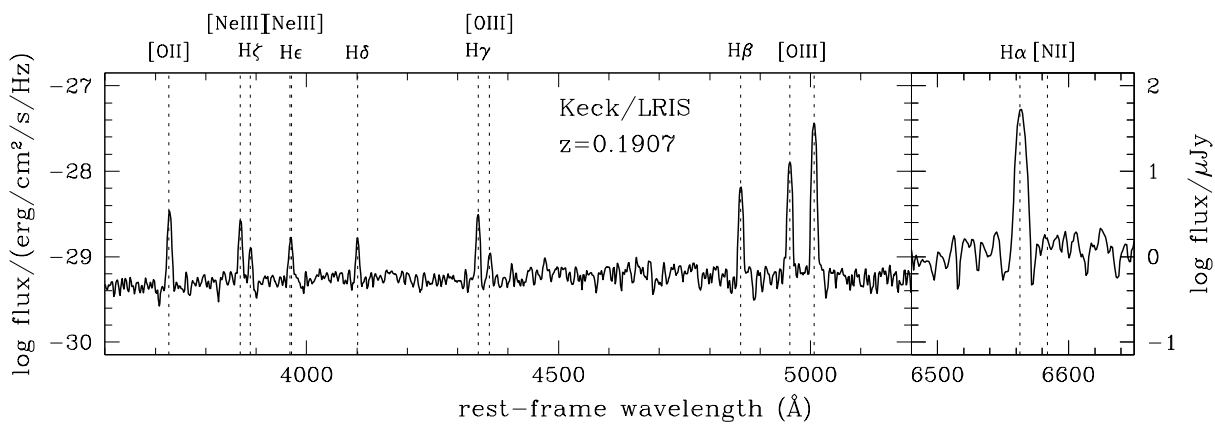}
\caption{The Keck/LRIS rest-frame spectrum (selected parts) of the transient's host galaxy with the major emission lines marked. 
From the line positions we find a redshift of $z=0.1907$. 
To derive the metallicity, we measured the fluxes of the [OII] 3727\AA, [OIII] 4363\AA, 
[OIII] 4959\AA, [OIII] 5007\AA\, and H$\beta$ 4861\AA\, lines.}                                                                                                                      
\label{fig:Keck_spectrum}                                                                                                   
\end{figure}

The Catalina Sky Survey (CSS, \citealt{2003DPS....35.3604L})/Catalina Real-Time Transient
Survey (CRTS, \citealt{2009ApJ...696..870D}) provides light curves for the period prior to epoch 2 and overlapping
epochs 3 and 4.  Each epoch of four 30 second exposures on the  0.7 meter Catalina Schmidt Telescope 
reaches an r-band magnitude limit of $\sim 20$~mag.   These data provide the strongest limit
on the optical-to-mid-IR flux ratios, where a stack of 12 images (3 epochs), gives a $3\sigma$ 
upper limit on the transient of $r>20.93$ mag, corresponding to $7.6\times 10^{42}$ ergs/s 
(large dark green symbol to the right in Figure~\ref{fig:lightcurve}), in the period between
epochs 3 and 4.

The QUEST (\citealt{2008AN....329..263D}) survey further limits the optical emission before epoch
2 and after epoch 4.
There are 19 epochs each consisting of 1 to 4, 60-100~sec exposures with the 1.2m Samuel Oschin
telescope in a wide red/IR filter that we calibrated to $i$-band.  The most interesting data were 
taken $\sim40$ days prior epoch 2, where we stacked six adjacent QUEST epochs (${\rm JD'} = 4276.71$--$4286.76$) 
to obtain $3\sigma$ detection limit of $i>20.70$ mag ($7.7\times 10^{42}$ ergs/s, large orange symbol in 
Figure~\ref{fig:lightcurve}).

Similarly, by coadding the high cadence RAPTOR (RAPid Telescopes for Optical Response, \citealt{2009ApJ...691..495W}) 
observations of the field, each of which is a 30~sec exposure using the RAPTOR-P array of four 200-mm Canon 
telephoto lenses, we obtained 15 upper limits at various epochs between 2004 and 2008, before and after the 
transient (Figure~\ref{fig:lightcurve}).  Calibrating the unfiltered observations to an $R$-band 
equivalent using Tycho 2 stars, we find no evidence for the transient, with $C_R > 18$~mag ($3\sigma$).

Only the northern station of the All Sky Automated Survey (ASAS) data (\citealt{1997AcA....47..467P}) obtained data during the
peak of the transient.  Combining 7 high quality ASAS images during the transient 
(${\rm JD'} = 4322.75$--$4525.11$), we obtain a weak upper limit
on the transient magnitude  of $V>16.3$ mag ($3\sigma$), corresponding to 
$6.2\times10^{44}$ ergs/s.  

\section{Discussion of the Transient }
\label{sec:results}

\begin{figure}[t]
\centering
\includegraphics[width=16cm]{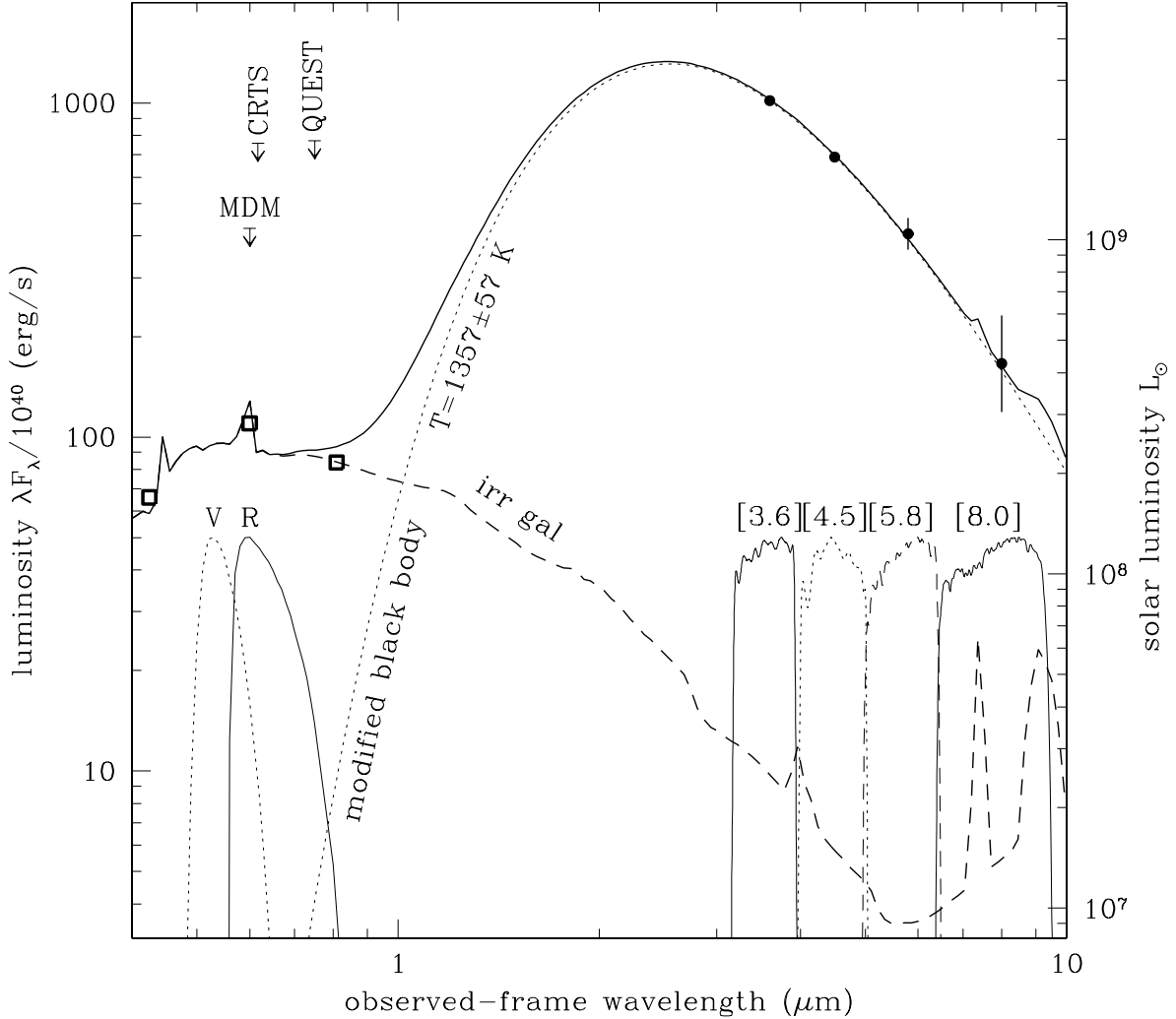}
\caption{SED of the transient during the second epoch of SDWFS. 
We fit a black body SED that includes the dust emissivity ($Q\propto\lambda^{-1}$) to the four IRAC bands (filled circles) and find
a best fitting temperature of $T=1357\pm57$ K (dotted line). We also show a template SED for the host galaxy modeled as an irregular
(dashed line; \citealt{2010ApJ...713..970A}) and the combination of the modified black body and irregular galaxy SEDs (thick solid line). 
Overplotted are the V, R and IRAC filter transmission curves, 
``normalized'' to $5\times 10^{41}$ ergs/s. The MDM (QUEST) limit is derived for an observation taken 90 ($\sim40$) days prior to the second epoch of SDWFS.
The CRTS limit corresponds to the stack of 3 epochs ($12\times30$ sec images) taken between SDWFS epochs 3 and 4.
The open squares mark the NDWFS $B_{\rm W}$, $R$, and $I$-band measurements for the host galaxy (from left to right).}                                                                                                                      
\label{fig:transientSED}                                                                                                   
\end{figure}

The steeply falling, roughly power law SED of the transient (Figure~\ref{fig:transientSED}) is suggestive
of thermal emission by a source cooler than normal stars.
If we fit the SED as dust emission using an emissivity of $Q\propto\lambda^{-\beta}$, we find good fits
with $\beta \simeq 1$ as is typical of the mid-IR emission by warm dust (e.g. \citealt{1980Ap&SS..72...79S, 1981ApJ...245..880D, 1986MNRAS.221..789G}).
The temperature of  $T\approx1350$~K varies little between epochs (Table~\ref{tab:SDWFSdata})
and is below typical dust destruction temperatures of $T\gtrsim1600$ K (e.g., \citealt{1983ApJ...274..175D,1989ApJ...345..230G}).  
Table~\ref{tab:SDWFSdata} summarizes these estimates of the temperatures and luminosities.
A zeroth order estimate (ignoring the dust emissivity) of the radius $R$ of the dust photosphere is
\begin{eqnarray}
\label{eq:radius}
  R &=& 8000   \left(\frac{L}{10^{43}~\rm ergs/s}\right)^{1/2}\left(\frac{1000~\rm K}{T}\right)^2 \rm~AU.
\end{eqnarray}
We are assuming that the shell is optically thick and $R$ corresponds to the radius where the (optical)
optical depth is of order unity.  If the dust is optically thin, as the case for typical dust echos,
there is no characteristic radius or temperature since photons can be absorbed and emitted at any 
distance from the SN.  We discuss these issues in more detail below.
For the second and third epochs of SDWFS, with total luminosity $L\approx6\times 10^{43}$ ergs/s
and temperature $T\approx1350$~K, we find the radius to be $R\approx11000$ AU (0.05 pc). 
The total energy radiated in the IR is then $E_{\rm rad} \approx 10^{51}$ ergs.  The weak optical
limits mean that a comparable amount of energy might have been radiated in the optical.

We used the more sophisticated radiation transport model DUSTY (\citealt{1997MNRAS.287..799I}) to check this estimate.
For incident black body spectra with temperatures 10,000--20,000~K, shell
geometries with constant density, and dust temperatures at the inner photosphere edge of 1300--2000~K,
we could match the epoch 2 and 3 SEDs, including a weak optical limit, using shells dominated by graphite,
a visual optical depth of $\tau\sim1$--3, and a radius $R\sim10,000$--$40,000$~AU.
Pure silicate dusts were unable to fit the data well, producing the wrong spectral shape for reasonable
$\tau$ and inner dust temperatures. These results are consistent with our simple estimates and
strongly suggest that there was an optical counterpart, albeit with significant extinction ($R \gtrsim 21.7$ mag).

Before outlining our preferred explanation of the transient as a self-obscured supernova, we briefly outline the 
hypotheses we rejected.

\subsection{Rejected Hypotheses}
\label{sec:results:rejectedhypo}

First, the transient cannot be explained by a Galactic star. The arguments against are as follows: 
(1) based on the Keck spectrum, we know that the source is spatially coincident with an irregular galaxy;
(2) few stars can brighten by at least 5 magnitudes (by more than a hundred times) 
    in the mid-IR and then stay at this level for 6 months before fading; and
(3) most flaring stars do not flare in the mid-IR but in the UV.  There are exceptions to
(2) and (3), and these will generally be accreting systems where the disk can alter the
spectral properties.  However, the NDWFS field is a high latitude ($b \simeq 67^\circ$) field with
negligible Galactic extinction and unlikely to contain young (protostars) or massive (Be stars) 
with these characteristics, particularly
if they must also have no optical counterpart in quiescence despite needing to be nearby
sources because of the short path length out of the disk.

The second possibility is that we are really observing an AGN.  First, we note that there are no
signs of AGN activity in the host spectrum.  The narrow lines in a galaxy are on relatively large
scales, and should show line ratios averaged over decades (or more) of AGN activity, so the lack
of a spectrum during the transient does not invalidate using them to argue against an AGN.  
Moreover, the mid-IR colors at the peak of the transient, $[3.6]-[4.5]=0.3$ mag and $[5.8]-[8.0]=0.0$ mag 
(epoch 2), are inconsistent with that of an AGN (e.g., \citealt{2005ApJ...631..163S, 2010ApJ...713..970A, 2010ApJ...716..530K}).
While the composite SED of an AGN and its host can have these colors, we know from the host flux
limits in the first epoch that the SED at the peak should be dominated by the AGN, and a pure 
AGN at $z\simeq0.2$ should have mid-IR colors closer to $[3.6]-[4.5] \simeq 0.8$ mag 
and $[5.8]-[8.0] \simeq 1.1$ mag (\citealt{2010ApJ...713..970A}).
We also considered the possibility of a tidally disrupted star in the context of the super-Eddington
accretion models of \cite{1997ApJ...489..573L}.  While these models have an inflated, relatively
cool envelope around the black hole, the expected temperatures of $T \approx 7000 (M_{\rm BH}/10^5$ M$_\odot)^{1/4}$~K
for a black hole of mass $M_{\rm BH}$ are still too high.  A black hole with low enough mass 
($M_{\rm BH}\approx150$ M$_\odot$) to match the temperature ($T \approx 1350$~K) would be unable to produce the observed luminosity. 
The Eddington luminosity for such a low mass is only $L_{\rm Edd}\approx2\times10^{40}$~ergs/s.

The third possibility is that it is a $\gamma$-ray burst (GRB).  We can rule it out as direct
emission from a burst because the time scale (6 months) is too long and 
the SED is wrong for direct beamed emission (e.g., \citealt{1999ApJ...519L..17S}).   It cannot be indirect
emission, where a dusty region absorbs and reradiates emission from the jet, because a
relativistic jet expands past the scale of the photosphere ($\sim 10^4$~AU) too quickly
($\sim 60$ rest-frame days) to produce the observed event duration. 

Finally, this cannot be gravitational lensing of a background object by the host galaxy 
(see \citealt{2006glsw.book..453W} for a review).  The
time scales are such that it would have to be microlensing by the stars in the lens galaxy,
but there are no known mid-IR sources that are sufficiently compact to show such a large,
and apparently achromatic, degree of magnification.  Moreover, almost any potential source
would have to show a still stronger degree of optical magnification that would disagree
even with our weak optical limits (see e.g., \citealt{2010ApJ...712.1129M}).  

\subsection{An Obscured Supernova Explosion}
\label{sec:results:SN}

Our working hypothesis is that SDWFS-MT-1 is a supernova analogous to SN~2006gy,
an extremely luminous Type~IIn SN 
(\citealt{2007ApJ...659L..13O,2007ApJ...666.1116S,2007ApJ...671L..17S,2008ApJ...686..485S,2010ApJ...709..856S,2010AJ....139.2218M}).  
SN~2006gy peaked at $M_V\simeq-22$ mag, corresponding to a luminosity of $L\sim3\times10^{44}$ ergs/s, 
and it radiated of order $E_{\rm rad}>2\times10^{51}$ ergs in total (\citealt{2007ApJ...659L..13O,2010AJ....139.2218M}).
In a normal Type~II SN, a very small fraction ($\lesssim1$\%) of the available energy is radiated. 
The initial energy from the shock is converted into kinetic energy by
adiabatic expansion, leaving little to power the luminosity.  Thus, a supernova
ejecting  $M_{\rm ej} \sim 10M_\odot$ at velocities of $v_{\rm ej}\sim 4000$~km/s
contains enough kinetic energy, $E_{\rm kin} \simeq 2 \times 10^{51}$~ergs to power
the transient, but it is only radiated on time scales of order 10$^3$ years as the SN remnant develops.
The scenario we present was
proposed by \cite{2007ApJ...671L..17S} for SN~2006gy with further elaborations
in \cite{2008ApJ...686..485S} and \cite{2010AJ....139.2218M}.
 
Very massive stars ($\gtorder 50 M_\odot$) are known to undergo impulsive
events ejecting significant fractions of the stellar envelope.  The most famous example
is the eruption of $\eta$ Carina in 1837-1857 (see \citealt{1997ARA&A..35....1D}), but also
includes some of the so-called ``supernova impostors'' (e.g.  \citealt{2006ApJ...645L..45S,2007Natur.447..829P}). The
number and temporal distributions of these mass ejections relative to the final supernova is 
unknown, but there are several cases where one seems to have occurred shortly before the final
supernova (e.g., \citealt{2007ApJ...666.1116S}).  Pair instability SNe can also produce
a sequence of mass ejections just prior to the death of the star (see \citealt{2007Natur.450..390W}).  
Let us suppose that the progenitor of SDWFS-MT-1 was such a
massive star and had two such eruptions in its final phases, roughly 300 and 5
years before the final explosion, each of which ejected of order $M_{\rm shell} \simeq 10 M_\odot$. 
For a typical ejection velocity of $200$~km/s (e.g. \citealt{2006ApJ...645L..45S}),
these shells would lie roughly 13000 and 200~AU from the star when it exploded. 

The role of the first shell is to convert the kinetic energy of the explosion 
into radiation following \cite{2007ApJ...671L..17S}.  Roughly speaking, the collision between the ejected material and
the first shell can extract fraction $M_{\rm shell}/(M_{\rm ej}+M_{\rm shell})$ of the
kinetic energy and convert it to radiation.  The distance to the shell is
tuned so that the shock crossing time is comparable to the Thomson photon
diffusion time, thereby allowing the radiation to escape without significant
adiabatic losses.  This requires a total mass of approximately
$M =4\pi R^2 m_{\rm p} c/\sigma_{\rm T} v_{\rm ej}=2.6 R_{100}^2 v_{\rm ej,4000}^{-1} M_\odot$ where $R=100 R_{100}$~AU
and $v_{\rm ej}=4000 v_{\rm ej,4000}$~km/s.  We can get up to $20M_\odot$ either by
making $R \simeq 300$~AU or by allowing for adiabatic losses.
This then implies characteristic
time, energy, luminosity and temperature scales of $t=R/v_{\rm ej}=50 R_{100}/v_{\rm ej,4000}$~days,
$E=(1/4) M v_{\rm ej}^2 = 2 \times 10^{50} R_{100}^2 v_{\rm ej,4000}$~ergs, 
$E/t=5 \times 10^{43} v_{\rm ej,4000}^2 R_{100}$~ergs/s, and 
$T = (L/4\pi R^2 \sigma)^{1/4}=14000 v_{\rm ej,4000}^{1/2} R_{100}^{-1/4}$~K, respectively, where
we have used the estimate of the total mass and equally divided
it between the shell and the SN ejecta.  The
basic energetics, but not the radiation transport, are largely
confirmed by the simulations of \cite{2009MNRAS.394..595V}. 
This set our choice of incident temperature in the DUSTY models. Thus,
a shell of ejected material at $R \sim 300v_{\rm ej,4000}^{-1}$~AU can produce the
necessary energetics and luminosities but cannot match the observed time
scales (too short) or temperatures (too high).

The role of the second shell is to match the time and temperature scales.  \cite{2008ApJ...686..485S} and 
\cite{2010AJ....139.2218M} invoke a second $M_{\rm shell} \simeq 10M_\odot$ shell located at 
$R \simeq 1$~pc in order to explain the relatively bright, long-lived IR emission observed for SN~2006gy.  
Their shell is, however, optically thin to optical light ($\tau \simeq 0.02$) because the 
total infrared emission is only a small fraction of the total.  The optical depth of
a shell is of order $\tau \simeq 0.1 \kappa (M_{\rm shell}/10 M_\odot) (10^4 \hbox{AU}/R)^2$
for $\kappa$ in cm$^2$/g, which leads to an optically thin shell with $\tau_{\rm opt} \simeq 0.08$
for an $R\simeq 1$~pc shell radius and $\kappa_{\rm opt} \simeq 500$~cm$^2$/g for a radiation
temperature of order $10^4$~K (e.g., \citealt{2003A&A...410..611S}).   This opacity, $\kappa \simeq 500$~cm$^2$/g,
corresponds to a dust-to-gas ratio of approximately 1\%, which arguably may be high for a
low metallicity galaxy.  Note, however, that SN~2006gy is probably also associated with a low
metallicity galaxy (see \S3.4) and the shell producing its dust echo contains enough dust for
our scenario.  

To explain SDWFS-MT-1 we simply move the shell inward. At this point, however, we should
improve on Eqn.~(\ref{eq:radius}) with a simple model for an optically thick dust echo.  
Suppose that the dust lies in an optically
thick spherical shell of radius $R$ and is exposed to luminosity $L_{\rm p}$ for a period of 
time $t_{\rm p}$.  Because all the dust is exposed to the same luminosity, the dust temperature
is independent of position or time while irradiated.  The observer, however, sees the
luminosity pulse modulated by the light travel time $2R/c$ across the shell.  In this simple model,
the light curves are trapezoids with linear rises and falls connected by a plateau created by
the changing fraction of the shell illuminated by the outburst from the observer's perspective,
but the dust temperature is constant. Models
that can reproduce the drop in luminosity between epochs 3 and 4 have the pulse time
$t_{\rm p} < 2R/c$, with the simplest model having the luminosity plateau run from 
epoch 2 to epoch 3.  The plateau duration is $2R/c - t_{\rm p} = 180$~days while the rise and
falls last $t_{\rm p}$.  The luminosity
of the plateau is $2 \pi R c t_{\rm p} \sigma T^4$, which is smaller than $L_{\rm p} = 4 \pi R^2 \sigma T^4$
by the ratio $2 t_{\rm p}/R c \simeq 0.2$ between the light crossing time and the pulse length (if $t_{\rm p} > 2 R/c$ 
the plateau luminosity is $L_{\rm p}$).
The factor of $3$ drop in luminosity between epochs 3 and 4 is $L_4/L_3 = 1 - t_{34}/t_{\rm p} \simeq 1/3$
where $t_{34} \simeq 33$~days is the time between these epochs.  Thus, we have $t_{\rm p} \simeq 50$~days, and $R \simeq 19000$~AU or
$115$ light days.  With these time and distance scales we come close to matching the epoch 2 and 3 
luminosities given the observed temperatures,
with an estimated luminosity of $L_2 = L_3 = 5 \times 10^{43}$~ergs/s.  The true luminosity is
$L_{\rm p} \simeq 2 \times 10^{44}$~ergs/s, which is why the radius estimate is somewhat larger
than the estimates of Eqn.~(\ref{eq:radius}).
There is some tension between the plateau luminosity and the drop in luminosity between the
last two epochs, but the overall agreement is good given the crudeness of the model both for
the dust emission and the luminosity profiles.  

The mass required to produce an a shell with optical depth $\tau$ at this distance is   
   $M \simeq  \tau (\kappa/500\hbox{cm}^2/\hbox{g})(R/19000~\hbox{AU})^2 M_\odot$,
so there is little difficulty having an optically thick shell
for masses of order $1$-$10 M_\odot$ even if the dust-to-gas ratio needs to be significantly
reduced from $\sim 1\%$ because of the low metallicity of the galaxy.   While the dust  
temperature is close to that for destroying dust (e.g., \citealt{1981ApJ...248..138D}), the situation is
somewhat different from a normal supernova because the (Thomson) optically thick inner shell protects the
outer from any initial high luminosity bursts of hard radiation.  For the IR-radiation, 
$\kappa_{\rm IR} \simeq 10$~cm$^2$/g, and the shell is  optically thin to the mid-IR emission
for reasonable shell masses, so the dust emission streams freely to the observer.  Given
the estimated pulse duration, $t_{\rm p} \simeq 50$~days, and luminosity, $L_{\rm p} = 2\times 10^{44}$~ergs/s,
we can estimate that the inner shell is at $R \simeq 160$~AU, the velocity must be $v_{\rm ej} \simeq 6300$~km/s
and the mass involved must be $M \simeq 4 M_\odot$, where the rapid decline pushes these
values to small radii/masses and high velocities in order to keep $t_{\rm p}$ short.  Given the
crudeness of the model, it is remarkable how well all the observational features can be
matched using relatively sensible physical parameters.

\subsection{The Transient's Host Galaxy}
\label{sec:results:host}

\begin{figure}[t]
\centering
\includegraphics[width=14cm]{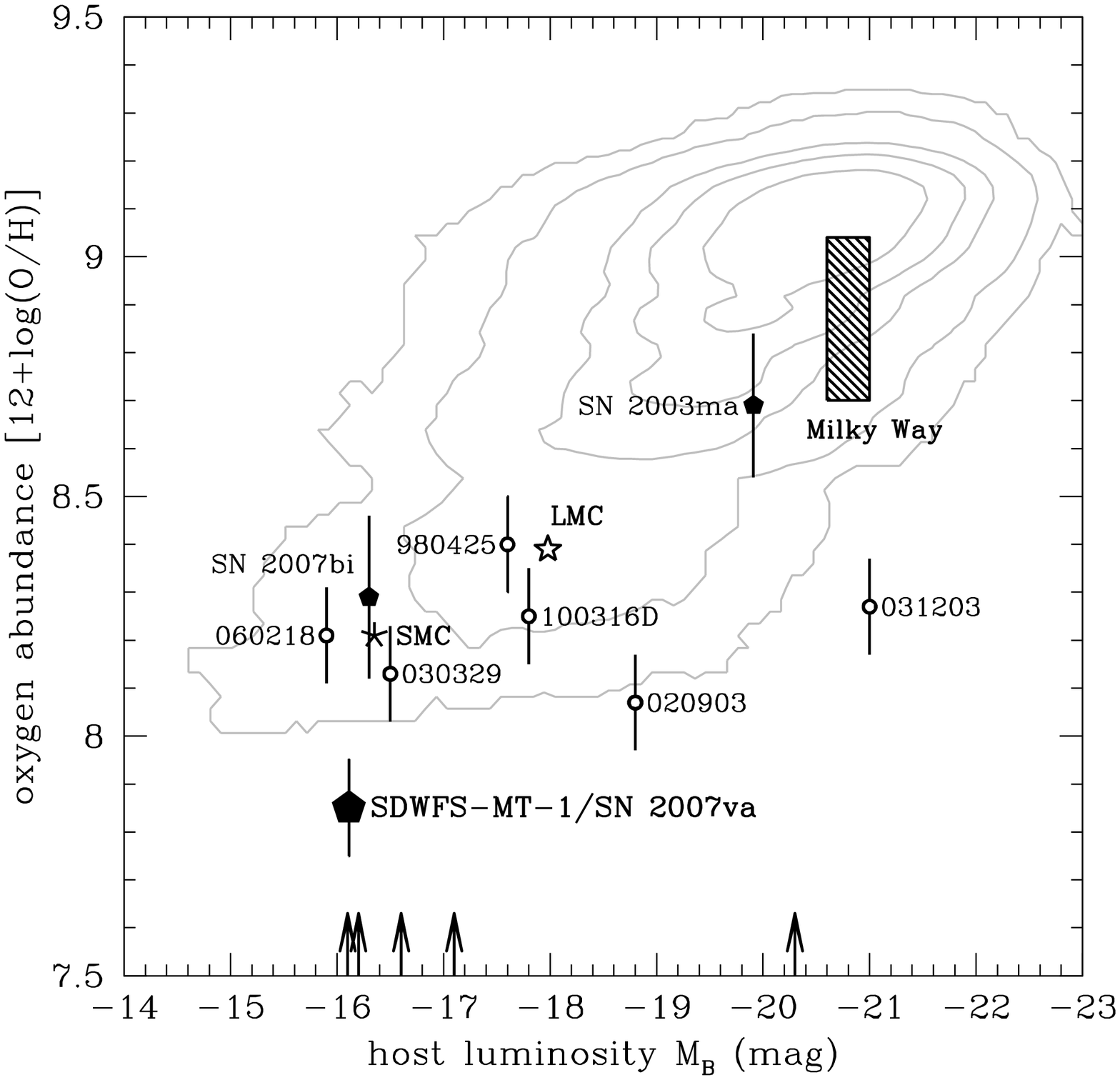}
\caption{Metallicity--luminosity relation for 125,958 SDSS DR4 star forming galaxies (e.g, \citealt{2004ApJ...613..898T, 2008ApJ...673..999P}). 
They are binned into 0.1 mag and 0.025 dex bins and
the smoothed contours ($3\times3$ bins) are drawn for 1, 10, 50, 100 and 200 objects per bin, counting from the outer contour.
The figure is analogous to Figure~1 from \cite{2006AcA....56..333S}. SDWFS-MT-1 is shown as the filled pentagon (lower left).
A handful of local GRBs with associated SNe (and two SNe) with known hosts metallicities and luminosities are shown with open circles (small pentagons).
The arrows show $M_B$ for five host galaxies of high luminosity SNe with unknown oxygen abundances. 
They are, from left to right, SN 2005ap, SN 2006tf, SN 2008fz, SN 2008es, and SN 2006gy
(see Section~\ref{sec:results:host}). The LMC/SMC oxygen
abundances are adopted following \cite{2008ApJ...685..904P} and Milky Way's (solar) are from \cite{2010arXiv1005.0423D}, 
while the absolute magnitudes are from \cite{2005AJ....129..178K}.}                                                                                                                      
\label{fig:Stanek}                                                                                                   
\end{figure}

The nature of the host galaxy indirectly supports this hypothesis because both GRBs and
most hyper-luminous Type IIn SNe are also associated with low luminosity, low metallicity
dwarf galaxies.  Figure~\ref{fig:Stanek}, similar to Figure~1 in \cite{2006AcA....56..333S}, shows the position of the host 
galaxy for SDWFS-MT-1 in the metallicity-luminosity plane in relation to  
$\sim126,000$ star-forming SDSS DR4 galaxies (\citealt{2004ApJ...613..898T,2006ApJS..162...38A,2008ApJ...673..999P}). 
We also include the hosts of six local ($z<0.25$) GRBs using updated estimates from
\cite{2010AJ....139..694L} and including the host of GRB 100316D (\citealt{2010arXiv1004.2262C,2010arXiv1004.2919S}).
While SDWFS-MT-1 was not a GRB, its occurrence in a very low metallicity galaxy is striking, 
since such galaxies produce only a small fraction of massive stars in the local universe 
(e.g., see Figure~2 in \citealt{2006AcA....56..333S}). In Figure~\ref{fig:Stanek}, we ignore small ($\lesssim20$\%) effects on metallicity measurements arising from the method
being used to derive it (see \citealt{2008ApJ...681.1183K}). 

In fact, the hosts of most other hyper-luminous SNe also tend to be very low luminosity 
($\sim 0.01\;L_{\star}$), suggestive of their low metallicity (unfortunately, oxygen abundances have 
not been measured for most of these hosts). SN~2005ap (\citealt{2007ApJ...668L..99Q}) at $z=0.283$ peaked at unfiltered magnitude 
of $M=-22.7$ mag and is associated with a $M_R=-16.8$ ($M_B\approx-16.1$) mag dwarf galaxy. 
SN~2006tf (\citealt{2007CBET..793....1Q,2008ApJ...686..467S}) 
at $z=0.074$ exploded in a host of low luminosity $M_r=-16.9$ ($M_B\approx-16.2$) mag. 
The over-luminous $M_R=-21.3$ mag Type 
Ic SN~2007bi exploded in a faint, $M_B=-16.3$ mag dwarf galaxy, with $12+{\rm log(O/H)} = 8.18 \pm 0.17$ 
(\citealt{2009Natur.462..624G,2010A&A...512A..70Y}). 
SN~2008es at $z=0.213$ (\citealt{2009ApJ...690.1303M,2009ApJ...690.1313G}) reached $M_V=-22.3$ mag and exploded in a host which was 
not detected down to a faint luminosity of $M_V>-17.4$ ($M_B>-17.1$) mag. 
SN~2008fz at $z=0.133$ peaked at $M_V=-22.3$ mag, 
but no host has been detected down to an upper limit of $M_R>-17$ ($M_B>-16.6$) mag (\citealt{2009arXiv0908.1990D}). 
SN~2006gy (\citealt{2006CBET..644....1Q,2007ApJ...666.1116S,2007ApJ...659L..13O}) is one of two exceptions here, as it 
exploded in a fairly luminous S0 galaxy, NGC 1260 with $M_B\approx-20.3$ mag.
Another example is SN~2003ma (\citealt{2009arXiv0911.2002R}) that released $4\times10^{51}$ ergs in a $M_B\approx-19.9$ mag, $12+{\rm log(O/H)}\approx8.7$ galaxy.

The general message is clear: GRBs and luminous SNe prefer low metallicity, 
low luminosity galaxies, so does SDWFS-MT-1.

\subsection{Rates}
\label{sec:results:rates}

We systematically searched for transients similar to SDWFS-MT-1 and SNe in general in the SDWFS data.
We considered highly correlated $[3.6]$ and $[4.5]$ light curves ($r>0.8$) with flag $=1$ (photometry unaffected by the wings of bright stars),
variability significances $\sigma_{[3.6]} >1 $ and $\sigma_{[4.5]}>1$ and colors $[3.6]-[4.5]<0.4$
to eliminate AGN (see \citealt{2010ApJ...716..530K}).  We selected as candidates objects that at their
peak were brighter than $[3.6]<18.67$~mag, corresponding to 1~mag brighter than the $3\sigma$ detection limit,
and which changed in brightness by at least one magnitude between either epochs 1 and 2 or epochs 2 and
3/4.  The latter two epochs are close enough together that an SN might not show a significant brightness
change between them.  We found six candidate objects, one of which is  SDWFS-MT-1.  
The next four of them are artifacts due to PSF structures (long spikes) from bright stars. 

The remaining object, SDWFS J142557.64+330732.6  (hereafter SDWFS-MT-2), peaks in epoch 2 close 
to our selection limit at $[3.6]=18.52\pm0.14$ mag and $[4.5]=18.71\pm0.33$ mag.
It lies $0\farcs7$ South-West of a potential host galaxy with {\it GALEX} $NUV=24.02\pm0.19$ mag, 
NDWFS $B_{\rm W} = 23.77\pm 0.03$ mag, $R = 23.34 \pm 0.04$ mag, and $I = 22.68 \pm 0.06$ mag that
is not detected in SDWFS.  In the deep images from NDWFS, the host galaxy appears to be an 
extended, possibly edge-on (length $3.0$ arcsec in $1.2$ arcsec seeing) disk, with a redder core and 
bluer outer regions. The transient lies in the bluer outer regions, although its position is
only known to $\sim 0\farcs2$ due to its faintness.  We estimate a photometric redshift of $z \approx 1.0$ 
for the host using the template models of \cite{2010ApJ...713..970A},
corresponding to  a comoving distance of $6.6$ Gpc and a linear scale of $8.0$ kpc/arcsec.
This translates into the linear diameter of the host of $\sim24$ kpc. The best fit model for
the SED of the host is a mixture of an AGN and a starburst template.  The mid-IR color
of the transient, $[3.6]-[4.5]=0.2\pm0.5$ mag, is ambiguous on the question of an SN 
or AGN origin for the transient, although the offset position would favor an SN. 
Given the redshift estimate, however, the implied luminosity is too high even for the
brightest normal SN  -- for $M_{[3.6]} \sim M_K \sim -20$ mag (\citealt{2003A&A...401..519M}) the peak
magnitude of $[3.6] \sim 24$~mag would be far below our detection limits.  A second event
like SWFS-MT-1 would have to be even more extreme.  Given the paucity of the data, interpreting
this source as a false positive seems the most probable explanation.

For events with six month durations like SDWFS-MT-1 our survey time coverage is approximately $\Delta T = 2$~years.
SDWFS-MT-1 peaked in epoch two at fluxes of  $0.12$ mJy in [3.6] and $0.10$ mJy in [4.5] at a luminosity distance
of $920$~Mpc.  The peak flux limit for our systematic search is $9.6\mu$Jy ($5\sigma$ in [4.5])
which means we could have detected SDWFS-MT-1 at a luminosity distance of $3000$~Mpc  ($920\times(100/9.6)^{1/2}$) or about $z=0.52$ 
ignoring $K$-corrections, which will initially aid detection because the SED peaks bluewards of the IRAC bands.
Thus, we could detect the source in a comoving volume of approximately $V=6 \times 10^6$~Mpc$^3$.  Combining
these factors, we estimate a rate for SDWFS-MT-1-like transients of $r = (V \Delta T)^{-1} \sim 8 \times 10^{-8}$~Mpc$^{-3}$~yr$^{-1}$
where we have not corrected the time scales for redshifting effects given the otherwise large uncertainties
from the statistics of one event.  This is approximately $10^{-3}$ of the Type II SNe rate (\citealt{2009PhRvD..79h3013H}), 
$\sim0.1$ of the hyper-luminous Type II SNe (\citealt{2009ApJ...690.1303M}), and $\sim10$ of the GRB rate (\citealt{2007PhR...442..166N}).  

\section{Summary}
\label{sec:summary}

In this paper, we presented the discovery of a luminous mid-IR transient, SDWFS-MT-1, that is unusual in several ways.
Its most striking features are the high luminosity, $L \sim 6 \times 10^{43}$~ergs/s, the low apparent temperature,
$T \sim 1350$~K, the long, 6 month duration of the event, and the resulting high amount of radiated energy, 
$E_{\rm rad}\gtrsim 10^{51}$ ergs. A model that explains the transient
is a self-obscured supernova, that has two concentric (patchy) shells, 
each with the mass of $\sim10$ M$_\odot$,  ejected prior to the SN explosion. The inner shell at $\sim 150$ AU
protects the dust in the outer shell from initial luminosity spikes and converts the high kinetic energy of the 
SN shock wave into thermalized radiation with $T\approx10^4$~K. The dust in the more distant (0.05 pc) second shell 
absorbs the radiation of the first shell, and reradiates it at $T\approx10^3$~K, while simultaneously smoothing
out and stretching the light curve. This scenario matches the observed energy, luminosity, time and temperature
scales of the transient by slightly modifying the scenarios used for other hyper-luminous Type IIn SNe 
(e.g., \citealt{2008ApJ...686..467S,2008ApJ...686..485S,2010AJ....139.2218M}). The host galaxy of the transient is also typical of these SNe, 
and our estimate of the rate is consistent with such events being $\sim10$\% of hyper-luminous Type II SNe.

The discovery of the progenitors of SN~2008S and NGC300-OT-1 (\citealt{2008ApJ...681L...9P,2009ApJ...705.1364T})
as self-obscured ($T\sim 400$~K, $L \sim 10^5 L_\odot$)  stars already indicated that mid-IR observations are
necessary to understanding the fates of massive stars.  If our interpretation
of SDWFS-MT-1 is correct, then this may extend to the actual transients as well as to their progenitors, although
it is quite likely that SDWFS-MT-1 was also a significant optical transient.  
IR dust echos from SNe are relatively common, (e.g., \citealt{1983ApJ...274..175D,2009arXiv0911.2002R, 2010AJ....139.2218M}),
but in these cases the optical depth of the dust is low and the IR emission is only a fraction of
what is observed in the optical.  A few recent SNe with significant echo luminosities are 
SN 2007od (\citealt{2010ApJ...715..541A}),  
SN 2006gy (\citealt{2007ApJ...666.1116S,2010AJ....139.2218M}), 
SN 2004et (\citealt{2009ApJ...704..306K}), and
SN 2002hh (\citealt{2006ApJ...649..332M}). 
Obviously there is a strong selection effect against optically selecting SNe where the optical depth of the
surviving dust is near unity, so the true abundance of such sources can only be determined in near/mid-IR
searches.  There have been some searches made in the near-IR (e.g., \citealt{2003A&A...401..519M,2007A&A...462..927C}), 
principally focused on the problem of unassociated foreground absorption rather than dust associated with the progenitor star.  

Two important steps are to 
try and identify massive stars with such dense shells of ejected material, similar to the search for 
self-obscured stars in \cite{2009ApJ...705.1364T} and \cite{2010ApJ...715.1094K}, and to better characterize
the mid-IR emission of other hyper-luminous Type IIn SNe.  If the inner shell invoked to convert the
kinetic energy into radiation is dusty, the progenitor will appear as a rapidly cooling near/mid-IR source 
as the shell expands. Such shells will have temperatures of order the 400~K observed for SN~2008S or NGC300-OT-1
but should show the time variability known to be absent for the SN~2008S progenitor (see \citealt{2008ApJ...681L...9P}).
 For the dust to survive the SN explosion and produce a powerful mid-IR transient, it
must be considerably more distant and colder than the $T \sim 400$~K dust photosphere of the progenitors
of SN~2008S or NGC300-OT-1, with temperatures of order $100$~K and SEDs peaking at 20-30$\mu$m that will 
make them invisible to warm {\it Spitzer}.  In fact, the progenitors may well resemble a moderately colder 
version of $\eta$~Carina, which radiates most of its energy in the mid-IR and has an SED peaking at 
$\sim10\mu$m (\citealt{1987MNRAS.227..535R}).  The temperature of such a distant shell will evolve slowly because the time scale
for the expansion of the shell is now quite long.  Warm {\it Spitzer} will remain a valuable tool for 
identifying warm, inner shells, but a careful search for cold outer shells will need to wait
for the {\it James Webb Space Telescope} ({\it JWST}).  
 
The {\it Wide-Field Infrared Survey Explorer} ({\it WISE}, \citealt{2005SPIE.5899..262M,2010arXiv1008.0031W}), 
launched in December 2009, is currently carrying out
an all-sky survey at similar wavelengths to {\it Spitzer}, as a natural by-product of its design,
find many transients similar to SDWFS-MT-1.  With a $5\sigma$ sensitivity of $0.12$ mJy at 3.4~\micron\ 
and $0.16$ mJy at 4.7~\micron\ for a region with 8 exposures (\citealt{2005SPIE.5899..262M}), 
{\it WISE} has far less sensitivity than the SDWFS survey.  Crudely speaking the survey volume compared
to SDWFS scales as the ratio of the $(solid~angle)(sensitivity)^{-3/2}$ of the two surveys, so
$V_{\rm WISE}/V_{\rm SDWFS} \simeq (42000/8)(0.12 \hbox{mJy}/6.3\mu\hbox{Jy})^{-3/2} \sim 10^2$ and the
greater solid angle wins over the reduced sensitivity.  If {\it WISE} completes two full years of
observations it will also have 4 epochs, 1 epoch every 6 months, and so should detect a significant
number of these mid-IR transients ($\sim 10^2$) independent of their origin.  For sources closer to the poles
of the {\it WISE} scan pattern, {\it WISE} will be able to produce detailed light curves.  On longer time
scales, {\it JWST} will be able to identify and monitor such luminous transients at almost any 
redshift.

\acknowledgments

We thank the anonymous referee, whose comments helped
us to improve the manuscript.
This work is based on observations made with the \sst, 
which is operated by the Jet Propulsion Laboratory (JPL),
California Institute of the Technology (Caltech) under contract
with the National Aeronautics and Space Administration (NASA).
Support for this work was provided by NASA through award numbers 1310744 
(C.S.K. and S.K.), 1314516 (M.L.N.A.) issued by JPL/Caltech. 
C.S.K., K.Z.S., T.A.T., and S.K. are also supported by National Science Foundation (NSF) grant AST-0908816.
J.L.P. acknowledges support from NASA through
Hubble Fellowship grant HF-51261.01-A awarded by STScI, which is
operated by AURA, Inc. for NASA, under contract NAS~5-2655.
This work made use of images and/or data products provided by NDWFS
(Jannuzi and Dey 1999; Jannuzi et al. 2005; Dey et al. 2005). The NDWFS and the research of 
A.D and B.T.J. are supported by the National Optical Astronomy 
Observatory (NOAO). NOAO is operated by AURA, Inc., under a cooperative agreement with NSF.
The CRTS survey is supported by NSF under grants AST-0407448 and AST-0909182. 
The CSS survey is funded by NASA under Grant No. NNG05GF22G issued
through the Science Mission Directorate Near-Earth Objects Observations Program.

\clearpage

\clearpage

\clearpage

\clearpage

\clearpage

\clearpage

\begin{deluxetable}{l|cccc}                       
\tablecaption{SDWFS Mid-IR Data In The Observed Frame.\label{tab:SDWFSdata}}
\tablewidth{0pt}                                                 
\tablehead{                                                      
\colhead{Band (Units)} & 
\colhead{Epoch 1} & 
\colhead{Epoch 2} & 
\colhead{Epoch 3} &
\colhead{Epoch 4}                        
}                                                                             
\startdata
$[3.6]$ (mag) & $>19.67$ & $15.93 \pm 0.01$ & $16.09 \pm 0.01$ & $17.99 \pm 0.08$ \\
$[4.5]$ (mag) & $>18.73$ & $15.61 \pm 0.02$ & $15.76 \pm 0.03$ & $17.72 \pm 0.15$ \\
$[5.8]$ (mag) & $>16.33$ & $15.44 \pm 0.15$ & $15.74 \pm 0.21$ & $16.93 \pm 0.55$ \\
$[8.0]$ (mag) & $>15.67$ & $15.43 \pm 0.30$ & $15.09 \pm 0.26$ & $>15.67$\\
\hline
$\lambda F_{\lambda} ([3.6])$ ($10^{40}$ ergs/s) & $<32$  & $1017 \pm 18$ & $878 \pm  15$ & $153 \pm 11$ \\
$\lambda F_{\lambda} ([4.5])$ ($10^{40}$ ergs/s) & $<39$  & $690 \pm  16$ & $601 \pm 19$ & $99 \pm 14$ \\
$\lambda F_{\lambda} ([5.8])$ ($10^{40}$ ergs/s) & $<179$ & $406 \pm  42$ & $308 \pm 61$ & $103 \pm 55$ \\
$\lambda F_{\lambda} ([8.0])$ ($10^{40}$ ergs/s) & $<133$ & $166 \pm  47$ & $227 \pm 55$ & $<133$\\
\cline{1-5}
\\
\\
 & \multicolumn{4}{c}{Black body fit including dust emissivity ($Q\propto\lambda^{-1}$)}\\
& \multicolumn{4}{c}{$E_{\rm rad}>0.9\times10^{51}$ ergs }\\
\cline{1-5}
\\
Black Body Temp. $(K)$                    & ---       & $1357\pm57$        & $1338\pm71$        & $1400\pm353$\\
``Total'' $L$ ($10^{43}$ ergs/s)         & ---       & $\sim6.8$          & $\sim5.7$          & $\sim1.8$\\
Radius $R$ (AU)$^a$               & ---       & $\sim11000$        & $\sim10500$        & $\sim5500$\\
\enddata        
\tablecomments{The SDWFS magnitudes were measured in 4 arcsec
diameter aperture with aperture corrections to 24 arcsec.\\
$^a$Radius $R$ corresponds to the radius where the optical depth $\tau_{\rm optical}\approx1$.}
\end{deluxetable}    

\end{document}